# Chiral Bobbers and Skyrmions in Epitaxial FeGe/Si(111) Films

Adam S. Ahmed,[1] James Rowland,[1] Bryan D. Esser,[2,3] Sarah R. Dunsiger,[4,5] David W. McComb,[2,3] Mohit Randeria,[1*] and Roland K. Kawakami[1**]

[1] *Department of Physics, The Ohio State University, Columbus, OH 43210, USA*
[2] *Center for Electron Microscopy and Analysis, The Ohio State University, Columbus, OH 43210, USA*
[3] *Department of Materials Science and Engineering, The Ohio State University, Columbus, OH 43210, USA*
[4] *Center for Emergent Materials, The Ohio State University, Columbus, OH, 43210, USA*
[5] *Department of Physics, Simon Fraser University, Burnaby, British Columbia, Canada V5A 1S6*

**Abstract**

We report experimental and theoretical evidence for the formation of chiral bobbers — an interfacial topological spin texture —in FeGe films grown by molecular beam epitaxy (MBE). After establishing the presence of skyrmions in FeGe/Si(111) thin film samples through Lorentz transmission electron microscopy and topological Hall effect, we perform magnetization measurements that reveal an inverse relationship between film thickness and the slope of the susceptibility ($d\chi/dH$). We present evidence for the evolution as a function of film thickness, $L$, from a skyrmion phase for $L < L_D/2$ to a cone phase with chiral bobbers at the interface for $L > L_D/2$, where $L_D$ ~70 nm is the FeGe pitch length. We show using micromagnetic simulations that chiral bobbers, earlier predicted to be metastable, are in fact the stable ground state in the presence of an additional interfacial Rashba Dzyaloshinskii-Moriya interaction (DMI).

*email: randeria.1@osu.edu

**email: kawakami.15@osu.edu

Skyrmions are localized spin textures that exist in magnetic materials where spatial inversion symmetry is broken [1-8]. In such systems, the Dzyaloshinskii-Moriya interaction (DMI) favoring perpendicular alignment of neighboring spins competes with the ferromagnetic exchange interaction and magnetic anisotropy to form a variety of non-collinear and non-coplanar spin textures including skyrmion, helical, and conical phases, whose stability depends on the external magnetic field ($H$) and temperature ($T$). The most well-studied are the non-centrosymmetric B20 crystals such as MnSi [5,9] and FeGe [10], where the phase is restricted to a small pocket of the bulk $H$-$T$ phase diagram near the magnetic ordering transition. Interestingly, studies of FeGe thin films produced either by thinning bulk crystals or by epitaxial growth have shown that the skyrmion phase occupies a much larger region of the $H$-$T$ phase diagram [10-12]. This enhanced stability of skyrmions has motivated theoretical studies of magnetic phase diagrams and novel spin textures in thin films [13-18].

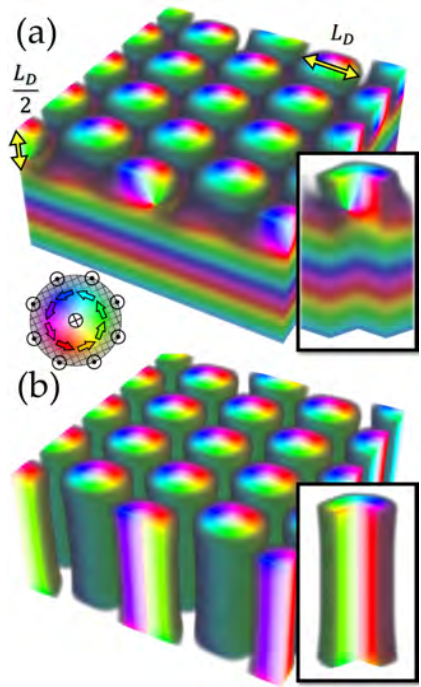

**Figure 1.** Examples of two different topological spin textures. (a) Chiral bobber crystal confined to the surface where the bulk is the conical phase. Inset: a 3/4 turn of a chiral bobber shown above the cone phase. (b) Skyrmion crystal. Inset: a 3/4 turn of a single skyrmion tube shown to extend through the entire sample. The color wheel indicates the orientation of the in-plane spins that turn in a counterclockwise fashion. The center point (white) indicates a spin into the page, and the grey/transparent perimeter are spins pointing out of the page.

One of the fascinating predictions for thin films of non-centrosymmetric materials is the presence of new spin textures forming at the surfaces and interfaces including the "chiral bobber" [19] and "stacked spiral" [20] phases. In Figure 1, we illustrate the interfacial chiral bobber crystal (Figure 1a) and compare with the well-known bulk skyrmion crystal (Figure 1b). The skyrmion phase consists of a hexagonal array of skyrmion tubes that extend throughout the crystal and are aligned with the external magnetic field. Each skyrmion tube consists of magnetic moments that wind about its centerline with a topological charge of one. Like skyrmions, chiral bobbers (Figure 1a) have moments that wind around a centerline and carry topological charge. However, unlike skyrmions, chiral bobbers are localized to the surface of a film in a region with thickness ~$L_D$/2, where $L_D$ is the helical pitch length, and terminate at a singular point (Bloch point). The remainder of the films is the topologically trivial cone phase. Until now, it has been unclear whether the chiral bobber phase could be realized because previous calculations [19] could only establish that it is a metastable state, not a true ground state.

In this Letter, we report experimental and theoretical evidence for a stable chiral bobber region through magnetization measurements on a series of epitaxial FeGe thin films grown by molecular beam epitaxy (MBE). After establishing the presence of skyrmions in our FeGe/Si(111) samples with thickness < $L_D$ through Lorentz transmission electron microscopy (LTEM) and topological Hall effect, we investigate the magnetic phase diagram through magnetization measurements ($M$ vs. $H$) and analysis of the susceptibility curves ($\chi$ vs. $H$) for different temperatures and film thicknesses ($L$). We provide evidence for interfacial chiral bobbers using a combination of experiment and theory. We show that the experimentally measured susceptibility has a slope ($d\chi/dH$) that is constant for $L < L_D/2$, and scales as $1/L$ for $L > L_D/2$. This implies an interfacial spin texture which penetrates a distance $L_D/2$ into the sample. We then use micromagnetic calculations to identify this spin texture as a skyrmion lattice in the very thin films and a chiral bobber lattice on the surface of a bulk cone phase in the thicker samples. We need to include two new ingredients – interface DMI and magnetic anisotropy – in the simulations to understand the experimental observations. It is known that the bulk (Dresselhaus) DMI of B20 materials leads only to metastable [19] chiral bobbers in the thin film geometry. We show that interface (Rashba) DMI, arising from broken surface inversion symmetry [13,18] in a thin film, together with the bulk DMI leads to stable interfacial chiral bobbers. Further, an analysis of our experimental saturation fields indicates an effective easy-plane magnetic anisotropy ($K_{eff}$) in our films. We show that this too is an important input parameter in the micromagnetic simulations that give us insight into the evolution from skyrmions to chiral bobbers with increasing film thickness.

Experiments are performed on FeGe/Si(111) films grown using MBE [21] and characterized using x-ray diffraction, atomic force microscopy, and cross-sectional TEM. Details of growth and characterization are found in the Supplemental Material [22] (additional references [23-26] found within).

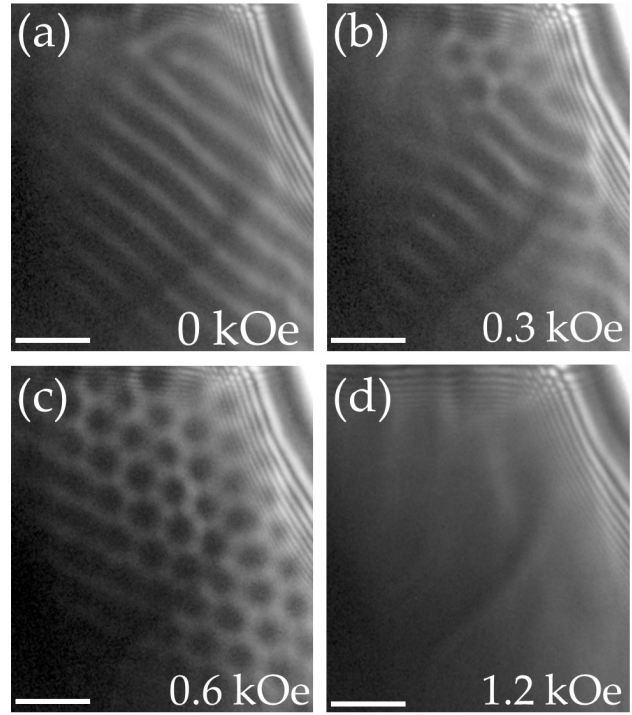

**Figure 2**. Series of LTEM images on a 55 nm thick cross section of an epitaxial FeGe film at 260 K. Both the [1$\bar{1}$0] crystal axis and the applied magnetic field point into the page. Different magnetic textures are shown for different field values. (a) 0 T: helical phase. (b) 0.3 kOe: coexistence of a skyrmion crystal and the helical phase. (c) 0.6 kOe: only skyrmion crystal present. (d) 1.2 kOe: field polarized state. Scale bars are 250 nm.

We perform LTEM measurements to demonstrate the presence of skyrmions in our FeGe films and establish the high quality of our material, and measure topological Hall effect to show consistency with previous experiments on sputtered FeGe films [11,12]. For the LTEM measurements, we extract a ~55 nm thick cross-section from an epitaxial 1 μm thick FeGe/Si(111) film, which takes advantage of standard TEM sample processing with focused ion beam milling. Figure 2 shows four representative LTEM images measured at $T = 260$ K with different magnetic fields $H$ applied along the transmission direction. For zero field (Figure 2a), the stripe pattern indicates the presence of the helical phase. For 0.3 kOe (Figure 2b), we observe the coexistence of the skyrmion phase (hexagonal lattice) and the helical phase (stripes). At 0.6 kOe (Figure 2c), the helical phase has disappeared, giving way to the skyrmion phase. Finally, for 1.2 kOe (Figure 2d), the disappearance of magnetic domain contrast signifies the field-polarized state. These LTEM images establish a qualitative baseline for the evolution of magnetic phases as a function of applied magnetic field along $[1\bar{1}0]$. We note that all other magnetic measurements (including topological Hall) have a different sample geometry with field along [111].

Next, we measure the topological Hall effect of a 35 nm FeGe film on Si(111), which is commonly used as a measure of "topological charge" density and can be suggestive of the presence of skyrmions. The topological Hall resistivity $\rho_{THE}$ is determined by etching the film into a Hall bar structure, measuring the Hall resistivity, and subtracting contributions from the ordinary Hall effect and anomalous Hall effects (details of the measurement and data analysis are provided in the Supplemental Material [22]). Figure 3a shows $\rho_{THE}$ as a function of applied field measured at $T = 50$ K. We observe a hysteresis by comparing the up sweep and down sweep, with a remanence of 42% at zero field. To obtain a $H$-$T$ mapping of $\rho_{THE}$, we measure the $\rho_{THE}$ vs. $H$ for a series of temperatures and plot $\rho_{THE}$ in the color plot of Figure 3b. In contrast to phase diagrams for bulk FeGe with small regions of skyrmion stability [27,28], the topological Hall effect occupies a wider range of the $H$-$T$ diagram. Notably, the observation of hysteresis and expanded $H$-$T$ range for $\rho_{THE}$ in our MBE-grown films is consistent with previous studies of $\rho_{THE}$ in sputter-deposited FeGe films [11,12].

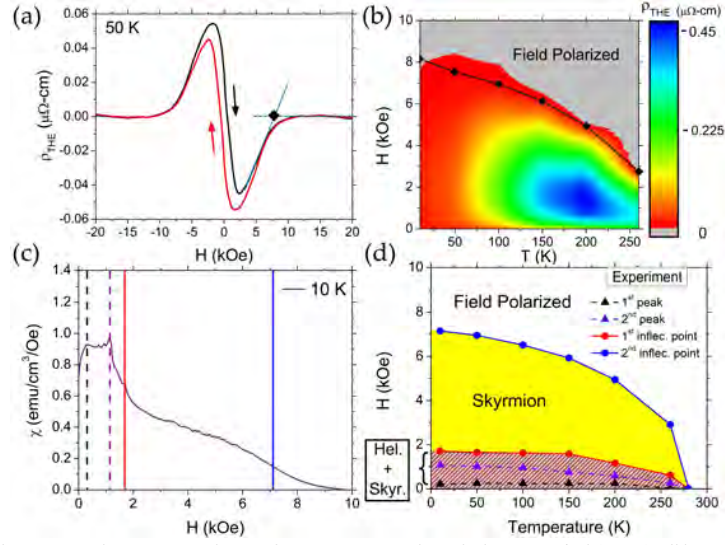

**Figure 3.** Electrical and magnetic properties of a 35 nm epitaxial FeGe/Si(111) film. (a) Topological Hall resistivity versus applied magnetic field at 50 K is hysteretic in applied field: up scan (black curve) and down scan (red curve). The black diamond represents the zero-crossing at high fields. The data was anti-symmetrized (see Supplemental Material [22]). (b) Average topological Hall resistivity as a function of field and temperature (10 K, 50 K, 100 K, 150 K, 200 K, 260 K). The black line (comprised of the extrapolated THE zero-crossings) represents the boundary between region of topological density and the field polarized regime. (c) Susceptibility obtained by numerical differentiation of *M* vs. *H* curves. SQUID scans are performed after cooling in zero field and measuring with increasing field steps. Solid lines indicate inflection points and dashed lines indicate peaks. (d) Magnetic phase diagram determined by features in the susceptibility data.

To search for novel magnetic phases in the FeGe films, we follow the methodology established by Bauer *et al.* [29,30] and perform magnetization measurements as a function of magnetic field, temperature, and film thickness. For all FeGe/Si(111) thin films, we employ a zero-field cool protocol to the desired temperature and measure *M* vs. *H* ramping from $H = 0$ Oe to 20,000 Oe in a superconducting quantum interference device (SQUID) magnetometer. Next, taking the numerical derivative yields the susceptibility $\chi = dM/dH$ vs. *H*. Figure 3c shows $\chi$ vs. *H* for a 35 nm FeGe film. The inflection points of $\chi$ correspond to phase transitions into different magnetic states (indicated by solid lines), local maxima correspond to magnetic structures in a state of coexistence (indicated by dashed lines) [23], and the low field double peak structure is believed to be related to helical reorientation in our films [31] (see discussion in Supplemental Material [22]). We repeat this procedure for a series of temperatures and generate the *H-T* magnetic phase diagram shown in Figure 3d, where the solid lines represent the phase boundaries and the dashed lines are the local maxima of $\chi$. Based on the magnetic phases observed in LTEM (Figure 2) and the *H-T* diagram for the topological Hall effect (Figure 3b), we assign the three phases as "helical + skyrmion" for low fields, "skyrmion" for intermediate fields, and "field polarized" for high fields.

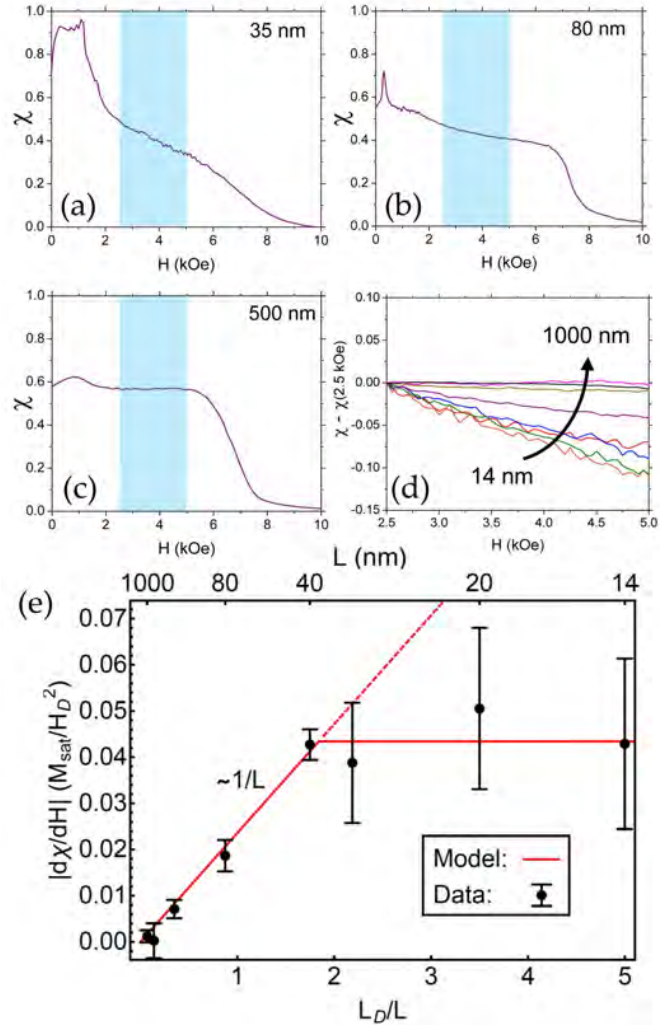

**Figure 4.** (a, b, c) Susceptibility versus magnetic field measured at 10 K for 35 nm, 80 nm, and 500 nm thickness of FeGe, respectively. The field is applied along the surface normal (growth direction). (d) Susceptibility, $\chi$-$\chi$(2.5 kOe), versus magnetic field taken at 10 K and plotted between [2.5 kOe, 5 kOe] for film thickness from 14 nm to 1000 nm. (e) Magnitude of $d\chi/dH$ plotted versus inverse thickness $L_D/L$, with $L_D$ = 70 nm. For thicknesses greater than 40 nm, $d\chi/dH$ follows a $1/L$ trend indicative of a surface phenomenon. For thicknesses less than 40 nm, $d\chi/dH$ deviates from the $1/L$ line (dashed red line) and is constant. Note: $M_{sat}$ is the saturation magnetization, and $H_D$ is the saturation field.

Exploring the thickness dependence of $\chi$ vs. $H$ identifies exciting new behavior. Figures 4a, 4b, and 4c show $\chi$ vs. $H$ for 35 nm, 80 nm, and 500 nm FeGe films, respectively. Looking to the 500 nm film in Figure 4c, there is a clear region of constant $\chi$ as function of field. The blue region between 2.5 kOe and 5.0 kOe represents the magnetic phase in this intermediate field range away from phase boundaries. We continue to track the evolution of this magnetic phase in this field range down to our thinnest samples. Comparing the three thicknesses in this field range, there is clearly a change in the slope of susceptibility, $d\chi/dH$. Figure 4d shows the $\chi$ vs. $H$ data in this field range for film thicknesses ranging from 14 nm to 1000 nm where the slope of $\chi$ clearly shows a thickness dependence. To understand the sign and thickness dependence of $d\chi/dH$, let us first focus on the thinnest films with $L$ much less than $L_D \sim 70$ nm [10,32,33] . The measured susceptibility is clearly decreasing as a function of $H$ in the same field regime where our micromagnetic simulations indicate a skyrmion state, and we observe a topological Hall effect indicative of topological spin textures (see, e.g., Fig. 3). Microscopically, our micromagnetic simulations show that negative sloped susceptibility is related to the reduction of the skyrmion core and chiral bobber core radius with increasing magnetic field (see the Supplemental Material [22]) which is substantiated by electron holography on thinned FeGe samples [34]. We thus use negative $d\chi/dH$ in the 35 nm film as indicative of a skyrmion state that evolves into a novel spin texture with increasing thickness.

From Figure 4e we see that there are two distinct regimes in the plot of the magnitude of the slope of the susceptibility, $|d\chi/dH|$, as a function of $1/L$. For small thicknesses ($L < L_D/2$), we find that $|d\chi/dH|$ is independent of film thickness $L$, while for thicker films ($L > L_D/2$) the slope magnitude scales like $1/L$. We next argue that such a scaling with $L$, with a crossover at $L_D/2$, is an unambiguous signature for an interface topological spin texture.

The illustration in Figure 1a depicts a film in the large $L$ regime, with chiral bobbers at the top interface. In this regime, the chiral bobber occupies a fixed volume (area × $L_D/2$) while the cone phase occupies the remainder of the sample of thickness $L$. As seen from Figure 1a, the chiral bobber is akin to a skyrmion at the surface, but which pinches off at a "Bloch point'' at a depth $L_D/2$ from the interface, as it merges with the conical texture in the interior. We note that the conical phase has a field-independent $\chi$, so it makes no contribution to $d\chi/dH$. Because $\chi$ is an intensive quantity (i.e., per unit volume), the contribution of chiral bobbers would scale like $\sim 1/L$, as seen in Figure 4e. On the other hand, for small thickness ($L < L_D/2$), the film can be thought of as "all interface" and there is no distinction between the chiral bobber and the skyrmion tube because both penetrate through the entire film. In this case, the contribution of the topological spin texture to $\chi$ is independent of film thickness. We note that another interfacial spin texture called stacked spiral [20] can also be stable in chiral magnetic systems, but we can rule out the possibility of stacked spirals forming in the region of negative $d\chi/dH$ as they are stabilized in low fields (see Figure 5b).

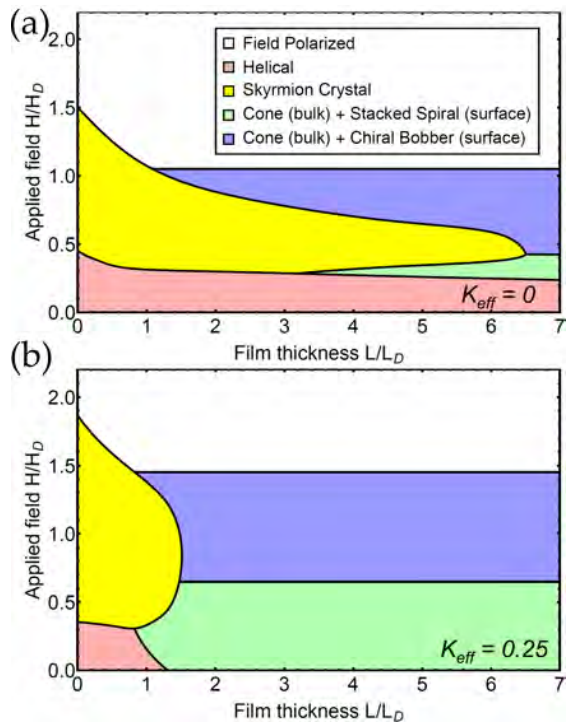

**Figure 5**. Applied field versus film thickness phase diagrams at $T = 0$ K calculated from micromagnetics. (a) Phase diagram with bulk and interface DMI and $K_{eff} = 0$ (no anisotropy). With sufficient interface DMI, the chiral bobber/cone phase becomes the true ground state instead of the pure cone phase. For $L/L_D \sim 1\text{-}6$, the chiral bobber/cone phase and the skyrmion phase have large regions of stability. (b) Phase diagram with bulk and interface DMI and $K_{eff} = 0.25$ (easy-plane anisotropy). The skyrmion phase is confined to low thicknesses, and the chiral bobber/cone phase dominates for $L/L_D \sim 1\text{-}6$. In addition, the stacked spiral/cone phase replaces the helical phase at low fields.

A final point to address is whether chiral bobbers can be stabilized. Previous calculations [19] show that with bulk DMI, chiral bobbers exist only as metastable textures on the surface of a cone phase, even in the presence of surface twists [35] that arise from free boundary conditions in thin films. However, an important consequence of the broken surface inversion symmetry in this geometry is the interface (Rashba) DMI, the effects of which have not been investigated heretofore. We present in Figure 5 the phase diagram based on micromagnetic simulations involving an energy functional that includes both interface and bulk

DMI, as required by symmetry for a B20 thin film on a substrate, in addition to the effects of surface twists [35]. We find that a chiral bobber crystal is now stabilized at the surface with interfacial DMI, while the rest of the system remains in a cone phase (Figure 1a); see Supplemental Material for details [22]. This establishes the theoretical basis for thermodynamically stable chiral bobbers.

We see from the zero-anisotropy phase diagram in Figure 5a that a system with thickness between $L/L_D$ = 1 to 6 would exhibit a transition from a skyrmion phase to chiral bobbers with increasing field. This is inconsistent with the experimental $\chi$, which shows a constant negative slope over the entire field range. From an analysis of our magnetization measurements, we find that our films have an effective *easy-plane* anisotropy (see Supplemental Material [22]). Upon including the effects of easy plane anisotropy in our simulations, we find (Figure 5b) that skyrmions are stable only for very thin films, while chiral bobbers are seen for large $L$, consistent with the experimental observations. We also see from Fig. 5b that the thick films that harbor chiral bobbers have upper and lower phase boundaries at $0.6*H_D$ = 2.2 kOe and $1.5*H_D$ = 5.5 kOe respectively (using the measured $H_D = M_s D^2/J$ = 3.7 kOe), consistent with our observations.

In conclusion, we have synthesized epitaxial FeGe thin films by MBE and investigated their magnetic phase diagram through LTEM, topological Hall effect, and magnetization measurements. Through systematic measurement of the thickness dependence of susceptibility, we observe a new interface-stabilized spin texture in thin films of non-centrosymmetric skyrmion materials. Micromagnetic simulations show that the presence of the chiral bobber phase stabilized by interface DMI and easy plane magnetic anisotropy explains both the thickness dependence as well as the magnetic phase diagram observed experimentally. To summarize, our combination of experiment and theory provides compelling evidence for the formation of chiral bobbers in FeGe thin films: the experimental data clearly shows the presence of an interfacial magnetic phase and the theoretical analysis identifies it as chiral bobber (+ cone phase bulk) with calculated phase boundaries (2.2 kOe, 5.5 kOe) that are consistent with the experimental field range. These results highlight the considerable potential for generating new magnetic phases in thin films and multilayers of B20 materials [21].

*Note added*. During preparation of this manuscript, we became aware of another study [36] which detected metastable chiral bobbers with a different experimental technique.

**Acknowledgements**

ASA and RKK acknowledge support from the Ohio State Materials Seed Grant (MTB-G00010). BDE and DWM acknowledge support from the Center for Emergent Materials at the Ohio State University, a National Science Foundation Materials Research Science and Engineering Center (Grant No. DMR-1420451), as well as the Ohio State Materials Seed Grant (MTB-G00012) and partial support from the Center for Electron Microscopy and Analysis. MR acknowledges support from NSF DMR-1410364. JR acknowledges support from the NSF graduate fellowship. SRD acknowledges support from the NSF MRSEC (Grant No. DMR-1420451). The Ohio State University Materials Research Seed Grant Program is funded by the Center for Emergent Materials, an NSF-MRSEC (DMR-1420451), the Center for Exploration of Novel Complex Materials (ENCOMM), and the Institute for Materials Research.

Supplemental Material for:

# Chiral Bobbers and Skyrmions in Epitaxial FeGe/Si(111) Films

Adam S. Ahmed,[1] James Rowland,[1] Bryan D. Esser,[2,3] Sarah R. Dunsiger,[4,5] David W. McComb,[2,3] Mohit Randeria,[1] and Roland K. Kawakami[1]

[1]*Department of Physics, The Ohio State University, Columbus, OH 43210, USA*
[2]*Center for Electron Microscopy and Analysis, The Ohio State University, Columbus, OH 43210, USA*
[3]*Department of Materials Science and Engineering, The Ohio State University, Columbus, OH 43210, USA*
[4]*Center for Emergent Materials, The Ohio State University, Columbus, OH, 43210, USA*
[5]*Department of Physics, Simon Fraser University, Burnaby, British Columbia, Canada V5A 1S6*

## 1. Materials growth and structural characterization

The Si(111) substrates were prepared beforehand with a chemical solvent sonication to remove any particulates or residue on the substrate surface (Acetone: 5 min., IPA: 5 min.). The Si(111) substrates were dipped in buffered HF solution, $NH_4F$ (33.0%) + HF (6.0%) + $H_2O$ (61.0%), for 2 minutes to remove the native oxide and hydrogen terminate the surface. This temporarily leaves the silicon surface resistant to oxidation for a few minutes. The substrates were rinsed with DI water and quickly loaded (air exposure time ~2 minutes) into the vacuum chamber.

Prior to deposition, the H-terminated Si(111) substrates were annealed at 800 °C for 20 minutes to 1) remove hydrogen from the surface and 2) convert the Si surface from a 1x1 to a 7x7 reconstructed surface. The 7x7 reconstructed Si surface is characteristic of well-ordered and atomically flat Si substrates. FeGe thin films were deposited on Si(111) substrates (MTI Corp.) with molecular beam epitaxy. The chamber base pressure before deposition was $3\times10^{-10}$ torr. Elemental sources of Fe and Ge were evaporated from thermal sources, and deposition rates were measured with a quartz crystal deposition monitor. The deposition rates were flux matched such that the ratio of Fe:Ge was 1:1. The substrate temperature was maintained at 300 °C to allow the FeGe thin films to crystallize in the B20 phase. Prior to deposition, a ~3 Å Fe seed layer was deposited to make an FeSi layer. FeGe thin films were then co-deposited until the desired thickness was achieved.

The samples were structurally characterized with *in situ* and *ex situ* methods. During the growth, the films were monitored with reflection high-energy electron diffraction (RHEED). This technique allows for real-time monitoring of in-plane diffraction information during growth. The samples were removed from vacuum and characterized with x-ray diffraction (XRD) to measure out-of-plane lattice spacings. Additionally, the samples were measured with cross-sectional scanning transmission electron microscopy (STEM) to image the crystal structure of the films and to characterize the interface quality between FeGe and Si.

We also employed Lorentz TEM to detect and image skyrmions in an epitaxial 1 μm thick FeGe film. A slice of the film was extracted with an FEI Helios NanoLab 600 DualBeam focused ion beam (FIB) with approximate dimensions of 50 μm wide, 10 μm tall, and a thickness of roughly 45 nm in the transmission direction. The magnetic field was applied parallel to the transmission direction using the objective lens. We confirmed the presence of skyrmions with this technique (Figure 2 of main text) and also extracted useful parameters such as skyrmion spacing and the in-plane components of the magnetic induction.

During growth, in-plane crystallinity was tracked with RHEED. Figures S1a and S1b show the RHEED images for a 40 nm FeGe film along the $[11\bar{2}]$ and $[1\bar{1}0]$ directions, respectively. The bright, sharp streaks are indicative of 2D terrace growth, and the presence of two different distinct crystallographic directions indicates the FeGe films are single crystal. For quantitative analysis of out-of-plane structural parameters, Figure S1c shows an XRD scan performed *ex situ*. The Si(111) and Si(222) peaks are present from the substrate. Additionally, an FeGe(111) peak is present, and no other phases of $Fe_xGe_y$ are present within our resolution. The FeGe(111) peak appears at a 2θ of 33.1° which corresponds to a layer spacing of 2.70 Å. This value agrees for previously measured FeGe films grown by MBE [1].

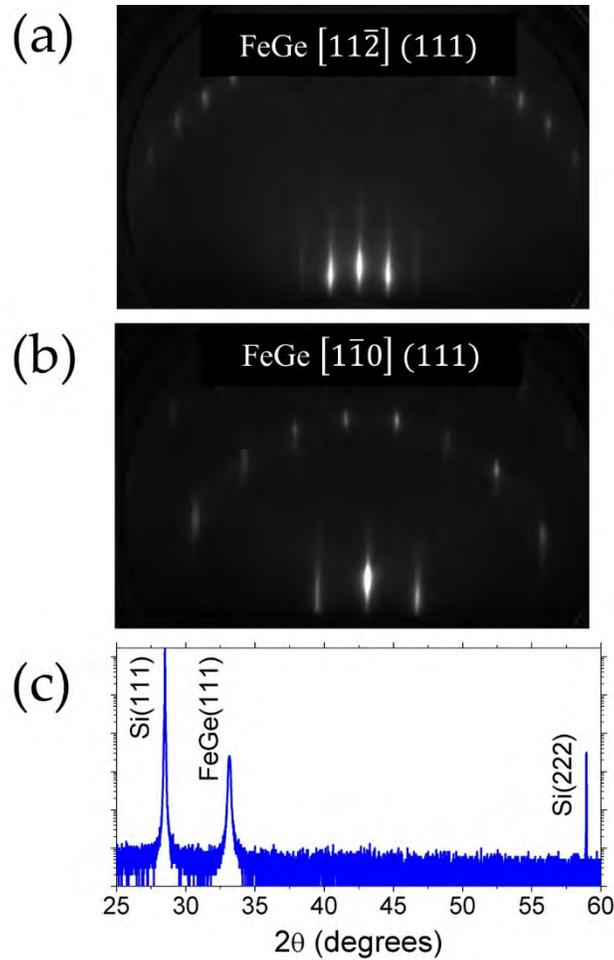

**Figure S1**. RHEED images on a 40 nm FeGe film along the (a) [11$\bar{2}$] and (b) [1$\bar{1}$0] directions. (b) XRD of 40 nm FeGe/Si(111) film. The FeGe(111) peak is present with no other phases present.

Lastly, cross-sectional STEM scans are shown in Figure S2 of a 1 μm FeGe film. The image shows single crystal structure of both FeGe and Si. The "zig-zag" structure seen in the FeGe film is characteristic for B20 crystals viewed along the [1$\bar{1}$0] direction [1]. The dark region between the FeGe film and Si is a transition region where Fe has reacted with the Si substrate and formed an interfacial B20 FeSi layer [1].

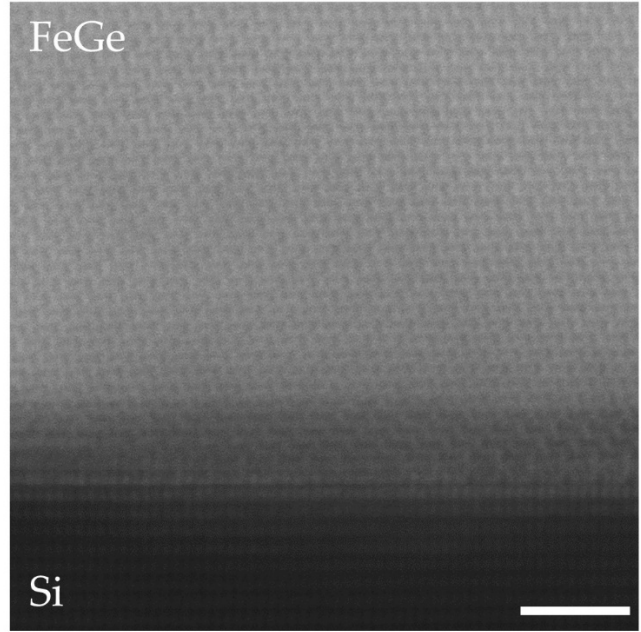

**Figure S2**. Cross-sectional STEM image of an FeGe/Si(111) film viewed along the [1$\bar{1}$0] direction. The image shows single crystal formation of FeGe. The transition layer is an FeSi layer after deposition of 1 ML of Fe/Si(111). The scale bar is 2 nm.

## 2. Topological Hall effect measurements and analysis

Hall bars were fabricated prior to electrical measurements. A photosensitive polymer (S1805) was spin coated onto the film, and Hall bar dimensions were defined using a laser writer. The exposed areas of the film were Ar-ion milled down to the substrate. Hall bars had a channel width of 0.5 mm and a channel length of 2 mm. The Si(111) substrates were undoped with resistivity > 10,000 μΩ-cm, so substrate shunting effects are negligible. Aluminum wires were wire bonded to the FeGe film directly, and the device was measured in a Quantum Design Physical Properties Measurement System (PPMS). The topological Hall component of resistivity is sensitive to spin textures with non-zero winding number—a characteristic of conduction electrons picking up a real-space Berry curvature as they traverse through skyrmions.

To detect the presence of spin textures with finite topological winding number, Figure S3a shows a 50 K scan of Hall resistivity for a 35 nm thin film. A current source was used to maintain a constant current of 200 μA, and 4-probe longitudinal and transverse Hall voltages were simultaneously measured. The total Hall resistivity, $\rho_{xy}$, is the sum of three contributions: the ordinary Hall effect ($\rho_{OHE}$, Lorentz force), the anomalous Hall effect ($\rho_{AHE}$, ferromagnetic background), and the topological Hall effect ($\rho_{THE}$, topological spin textures), i.e.,

$$\rho_{xy} = \rho_{OHE} + \rho_{AHE} + \rho_{THE}$$

$$= R_0 H + S_a \rho_{xx}^2 M + \rho_{THE}$$

where $R_0$ is the ordinary Hall coefficient and $S_a$ is a field independent parameter for the AHE. From the raw data shown in Figure S3a, we see the presence of a small hysteretic loop near zero field. This loop is characteristic of topological magnetic structures contributing to $\rho_{THE}$. However, there is a dc offset and the linear portion in high fields has slight curvature. These additional features are most likely due to longitudinal resistivity mixing with the transverse signal. To extract the pure Hall resistivity from the raw data, we take the anti-symmetric component because $\rho_{xy}$ must be anti-symmetric under time reversal symmetry. We use the equations

$$\tilde{\rho}_{xy}^+(H) = \frac{1}{2}\left[\rho_{xy}^+(H) - \rho_{xy}^-(-H)\right]$$

$$\tilde{\rho}_{xy}^-(H) = \frac{1}{2}\left[\rho_{xy}^-(H) - \rho_{xy}^+(-H)\right]$$

where $\tilde{\rho}_{xy}^+$ and $\tilde{\rho}_{xy}^-$ are the anti-symmetrized Hall resistivities, and $\rho_{xy}^+$ and $\rho_{xy}^-$ are the raw data for the Hall resistivity for increasing and decreasing magnetic field scans, respectively. The anti-symmetrized data are shown in Figure S3b. Henceforth, we refer to the anti-symmetrized data simply as $\rho_{xy}$.

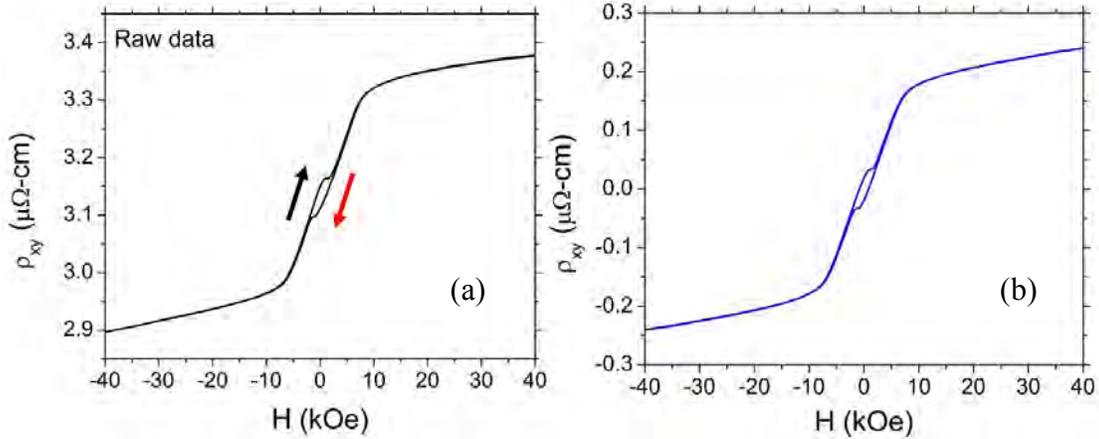

**Figure S3**. Transverse resistivity ($\rho_{xy}$) versus external magnetic field shown for 35 nm FeGe film at 50 K. Data shown for (a) raw data and (b) anti-symmetrized data.

To extract $\rho_{THE}$ from the Hall resistivity, we follow the protocol established in previous studies [2-5]. First, the out-of-plane M vs. H was measured by SQUID magnetometry (Figure S4a). Next, the OHE and AHE contributions were determined by fitting the high field saturated regime (H between 20 kOe and 50 kOe) as follows. In this field regime, there are no topological spin textures present and $\rho_{THE} = 0$, so the Hall resistivity is given by $(\rho_{xy}/H) = R_o + S_a \rho_{xx}^2 (M/H)$. By plotting $(\rho_{xy}/H)$ as a function of $(M/H)$, we determined $R_0$ (intercept) and $S_a \rho_{xx}^2$ (slope) by a linear fit. The results of the fit are shown in Figure S4b.

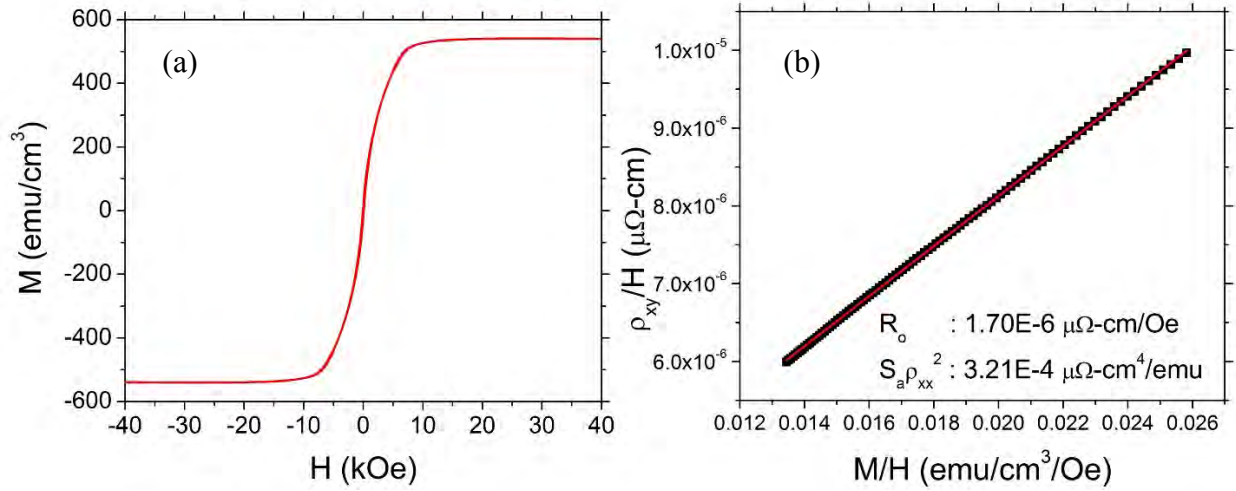

**Figure S4**. (a) Magnetization versus magnetic field. The field is applied out-of-plane. (b) Extraction method to obtain ordinary Hall and anomalous Hall coefficients $R_o$ and $S_a \rho_{xx}^2$, respectively. The magnetic field region is between 20 kOe and 40 kOe—well above magnetic saturation where $\rho_{THE}$ vanishes.

For our FeGe films, the longitudinal resistivity was tracked as a function of field and temperature. Figure S5a shows longitudinal magneto-resistance (MR) data defined as

$$MR\% = \frac{(\rho_{xx}(H) - \rho_{xx}(0))}{\rho_{xx}(0)} \times 100$$

From this we see that over the entire field range, the $\rho_{xx}(H)$ changes by less than 0.3% showing that $\rho_{xx}$ is approximately field independent. Therefore, the product of $\rho_{xx}^2$ and $S_a$ is also field independent. Additionally, we included the temperature dependence of $\rho_{xx}$. The decreasing resistivity with decreasing temperature implies that our FeGe epitaxial films are metallic with a residual resistivity ratio (RRR) of ~8, comparable to FeGe/Si(111) sputtered films.

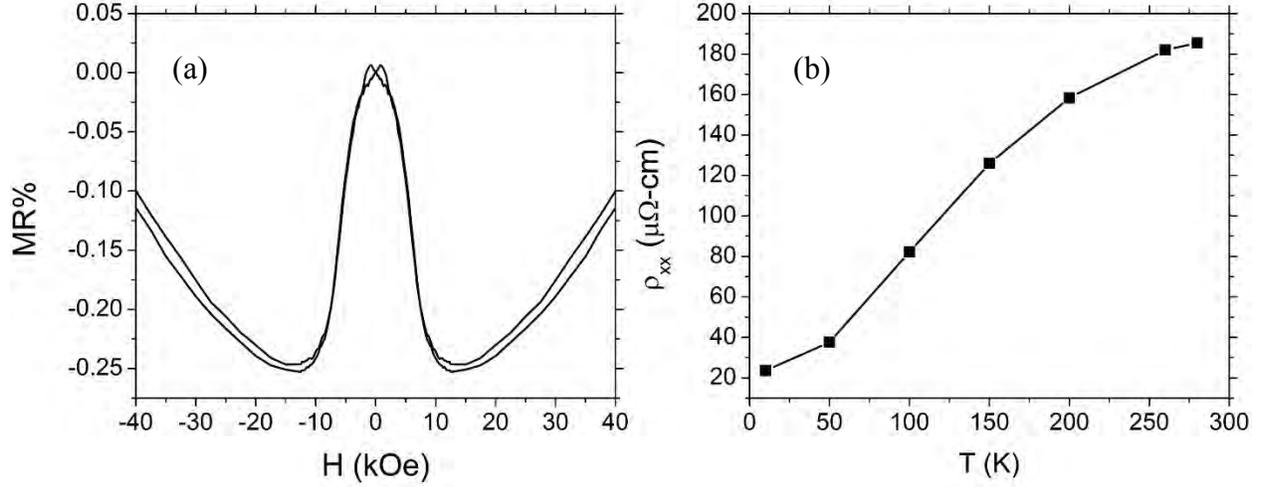

**Figure S5**. (a) Magneto-resistance data shown for 35 nm FeGe film taken at 50 K. MR% is less than 0.25% for all field values. (b) Longitudinal resistivity (at H = 0) versus temperature. The decreasing $\rho_{xx}$ with decreasing temperature shows that the FeGe films are metallic with an RRR of ~8.

After these procedures are complete, $\rho_{THE}$ is extracted by subtracting out $\rho_{OHE}$ and $\rho_{AHE}$ from $\rho_{xy}$. An example of a $\rho_{THE}$ curve is plotted in Figure S6. The hysteresis in $\rho_{THE}$ can be attributed to the meta-stability of skyrmions in zero field. A set of $\rho_{THE}$ curves was processed for several different temperatures, and a color plot of $\rho_{THE}$ as a function of *H* and *T* is shown in Figure S7. Data was taken from 10 K to 300 K in steps of approximately 50 K and smoothed over the entire temperature range. The data plotted in Figure S7 is for field decreasing from 50 kOe (saturation) to 0 kOe. In Figure S7, the largest THE signal occurs in a temperature regime of 150 K to 200 K in an applied field region of 0.5 kOe to 2.5 kOe. This phase diagram is similar to previous reports of epitaxial FeGe thin films [4,5] and also shows a large region of stability for skyrmions and/or topological spin textures in the MBE grown films.

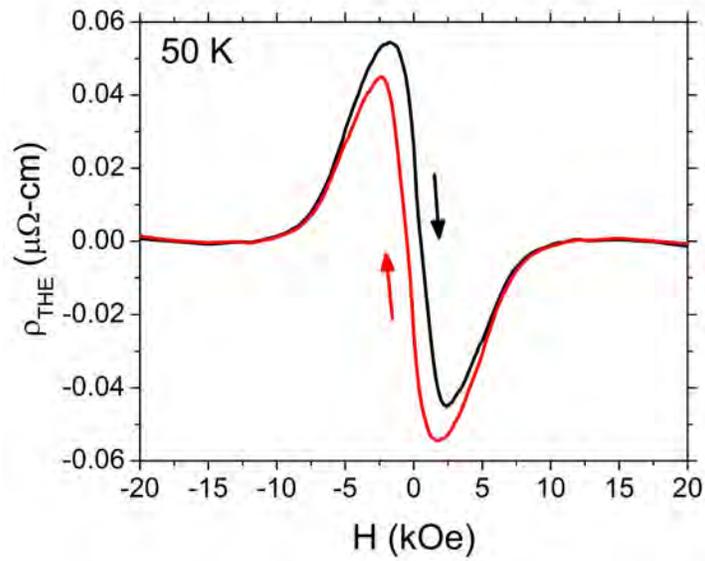

**Figure S6**. Topological Hall resistivity versus applied magnetic field for a 35 nm film at 50 K.

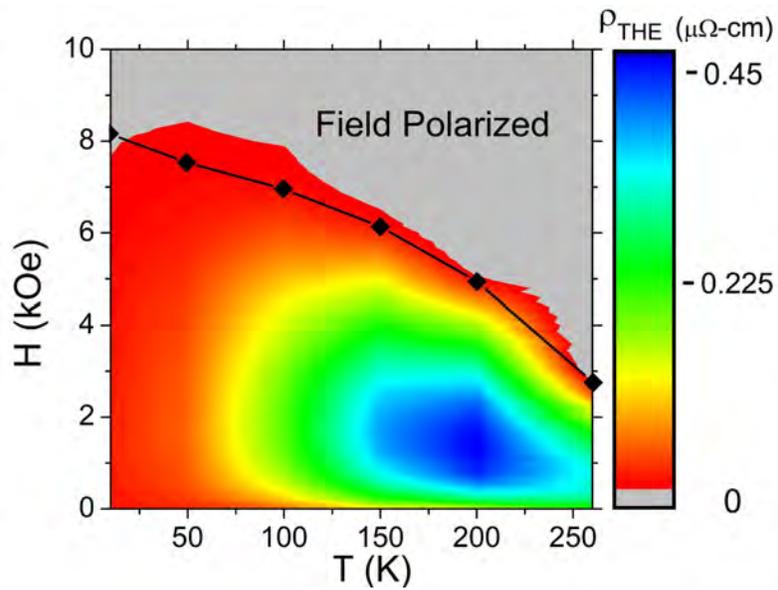

**Figure S7**. Topological Hall resistivity as a function of field and temperature (same as Figure 3b of the main text).

Lastly, we include line cuts for all temperatures in Figure S8 that form the contour plot shown in Figure S7.

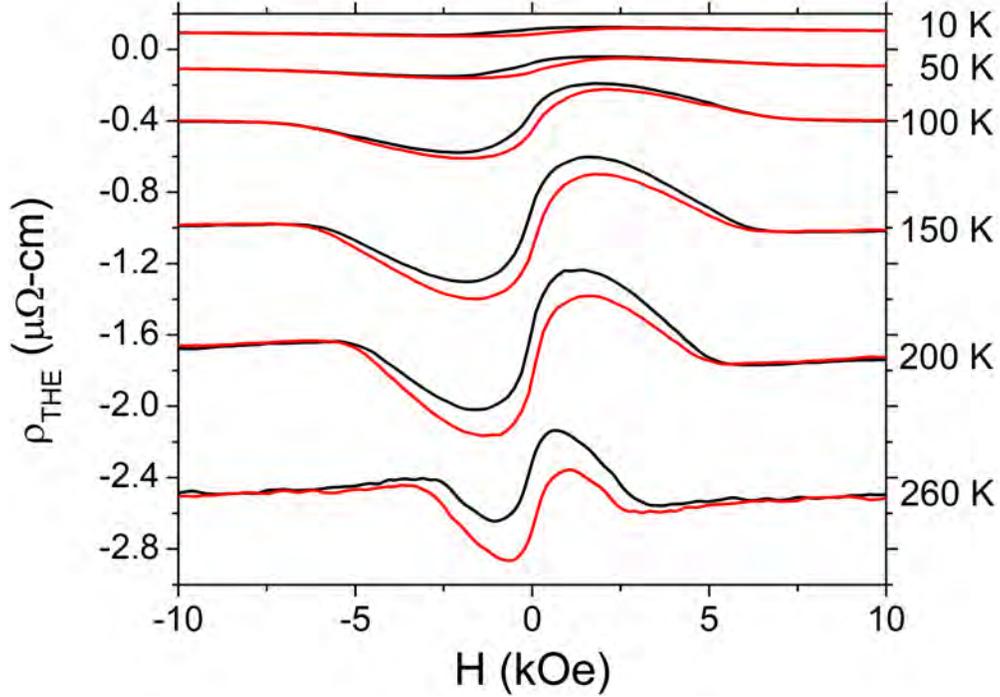

**Figure S8**. Topological Hall resistivity as a function of field and temperature.

### 3. Magnetization measurements for determining the magnetic phase boundaries

To determine the magnetic phase diagram of FeGe films in the *H-T* plane, we employ a method based on bulk magnetization measurements that has been successfully applied to MnSi, $Fe_{1-x}Co_xSi$, and $Cu_2OSeO_3$ [6-8]. The study by Bauer *et al.* used a combination of magnetization measurements (*M* vs. *H*) and ac susceptibility measurements ($\chi_{ac}$ vs. *H*) to identify phase boundaries in MnSi bulk crystals [6]. For the magnetization measurements, the susceptibility χ was obtained by numerical differentiation of the *M* vs. *H* curve (χ ≡ *dM/dH*). By comparing where *dM/dH* deviated and/or agreed with ac susceptibility measurements, a systematic method was developed for identifying features in *dM/dH* (e.g. peaks, inflection points) that correlate the responses to known phase diagrams for MnSi. Importantly, they found that *dM/dH* was more sensitive at revealing phase boundaries than ac susceptibility measurements, and peaks in *dM/dH* indicated mixed states (e.g. helix + cone) whereas inflection points represented phase boundaries [9]. In their bulk MnSi samples, the zero field ground state was determined to be the helical phase, and regions of constant susceptibility were characteristic of the cone phase and/or the skyrmion phase.

We applied these methods to map out the magnetic phase diagrams of the FeGe thin films. First, we found that ac susceptibility measurements on FeGe thin films had insufficient signal-to-noise due to the small net magnetic moment of the thin films. Therefore, we focused exclusively on the magnetization measurements, which were performed in a Quantum Design SQUID magnetometer using the following protocol. At room temperature (above the FeGe ordering temperature of 280 K), the magnetic coils were trained to remove background flux trapped in the coils. The FeGe/Si(111) films were zero-field cooled (ZFC) to a particular temperature, which avoided hysteretic effects and yielded the thermodynamic equilibrium state. Next, an $M$ vs. $H$ scan was measured at constant temperature by ramping the out-of-plane field from zero to 20 kOe, which is well above the field required to saturate the magnetization. Finally, the temperature was increased back to room temperature where the magnet coils were trained again to remove trapped flux within the coils, and this process was repeated for scans at different temperatures. Susceptibility was calculated from the $M$ vs. $H$ data as the numerical derivative ($\chi \equiv dM/dH$). By tracking the peaks and inflection points in the $\chi$ vs. $H$ curves obtained for different temperatures, we were able to determine the magnetic phase diagram in the $H$-$T$ plane.

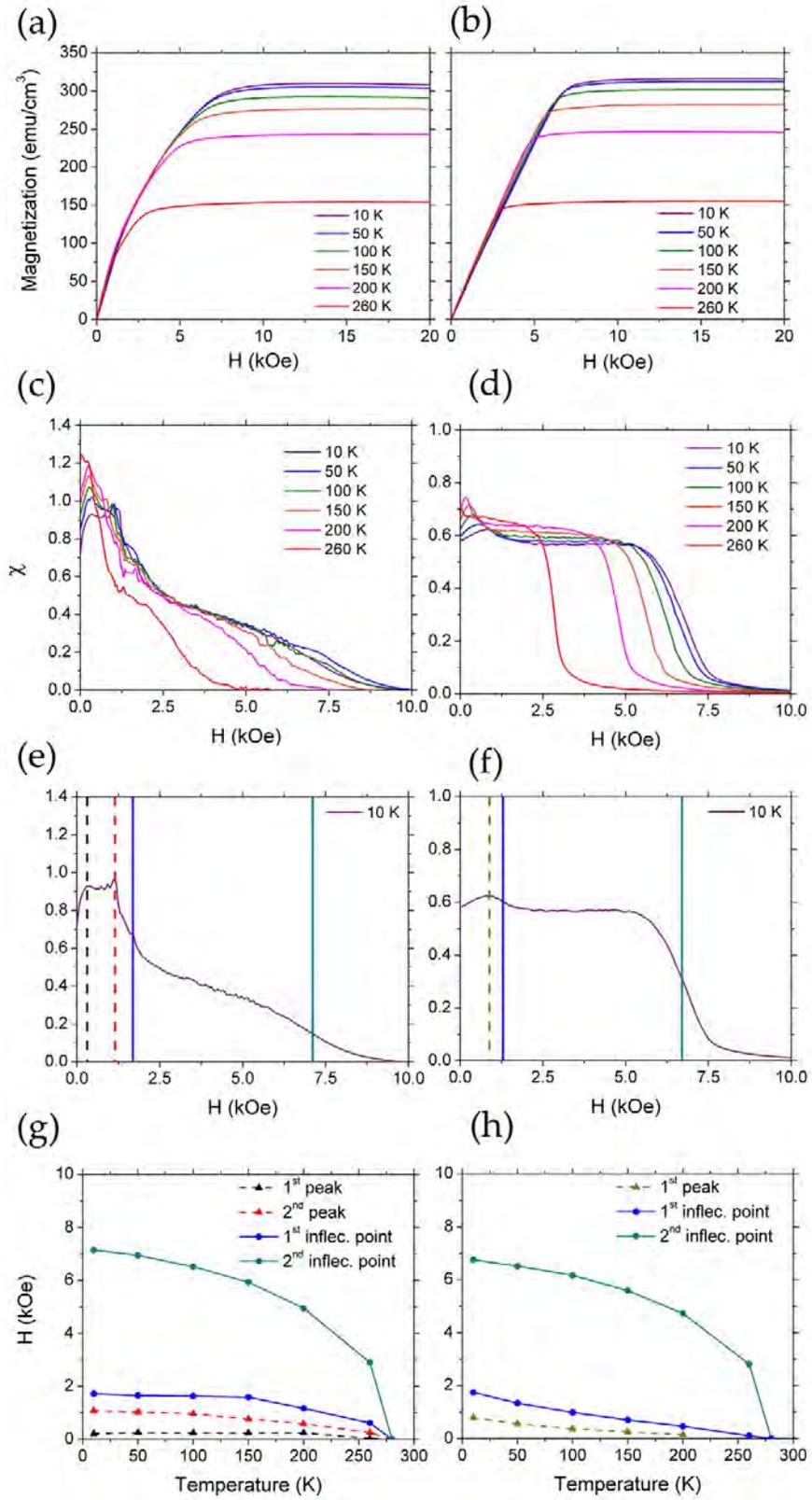

**Figure S9.** Magnetic data for a FeGe/Si(111) films for thicknesses of 35 nm (left column) and 500

nm (right column). (a, b) Magnetization versus applied field plotted for several temperatures below $T_c$. (c, d) Susceptibility versus applied field as the numerical derivative of $M$ vs. $H$ shown in (a) and (b). (e, f) Identification of peaks (dashed lines) and inflection points (solid lines) in $\chi$ vs. $H$ curves at $T = 10$ K. (g, h) Magnetic phase diagrams as a function of $H$ and $T$, obtained by tracking peaks and inflection points in $\chi$ vs. $H$ curves measured at various temperatures.

We present two representative cases in Figure S9 for an FeGe 35 nm film (left column) and an FeGe 500 nm film (right column). Figures S9a and S9b show magnetization versus field data for temperatures of 10 K, 50 K, 100 K, 150 K, 200 K, and 260 K. The saturation magnetization increases as temperature decreases, which is consistent with the temperature dependence for helimagnets. Below saturation, there is more curvature in $M$ vs. $H$ lines for the 35 nm film compared to a more linear trend in the 500 nm film. Susceptibility, $\chi$, is plotted in Figure S9c and S9d as the numerical derivative of the $M$ vs. $H$ data presented in Figures S9a and S9b. The data in Figures S9c and S9d have been smoothed. Naturally, noise is more prominent for the 35 nm film compared to the 500 nm film because thinner samples have a less magnetic moment and hence a smaller signal. As shown in Figure S9e, the $\chi$ vs. $H$ curve for a 35 nm film measured at 10 K shows a double peak behavior for low fields (< 2 kOe) and evolves into a negatively sloped susceptibility from ~2 kOe to ~6 kOe. Finally, when the film is field polarized (FP), the magnetization saturates and $\chi$ goes to zero. The peaks of susceptibility are indicated by dashed lines and inflection points of susceptibility are indicated by solid lines in Figure S9e. For comparison, the $\chi$ vs. $H$ curve for a 500 nm film measured at 10 K (Figure S9f) shows a single peak for low fields and a wide region of constant susceptibility from ~2 kOe to ~5 kOe. These features have been tracked and extracted from the data taken at different temperatures (Figures S9c and S9d) and compiled as phase diagrams in the $H$-$T$ plane as shown in Figures S9g and S9h for the 35 nm and 500 nm films, respectively. The dashed lines correspond to peak features that indicate mixed phases, while the solid lines correspond to inflection points that indicate phase boundaries. To elucidate some of the phase transitions shown in Figure S9e and S9f, we present second derivative data in Figure S10a for a 35 nm film and Figure S10b for a 500 nm film.

We now comment on the low field, double peak feature in $\chi$ in the 35 nm film. We believe the double peak feature in our FeGe films can be associated with helical reorientation where different helical domains change direction at different fields. This shallow double peak feature in susceptibility was previously shown to be present in bulk MnSi [10]. In MnSi, the helices preferentially point along the [111] family of magnetocrystalline axes. The relative angle between the external field and the [111] axes effects the dynamical motion to reorient the helices and manifests as a double peak in magnetic susceptibility. The abrupt motion of the helical q-vector orienting itself along the field direction is a result of different domains responding at different field values. We believe that the shallow double peak feature in our FeGe films also has this helical reorientation. There is precedence for helical reorientation in bulk FeGe as the magnetocrystalline axes are temperature dependent. Specifically, the preferred magnetocrystalline axis changes from being the (100) family of axes above 211 K to the [111] crystalline axes for temperatures less than 211 K upon cooling [11]. However, further studies are needed to elucidate the exact nature of the low field double peak feature.

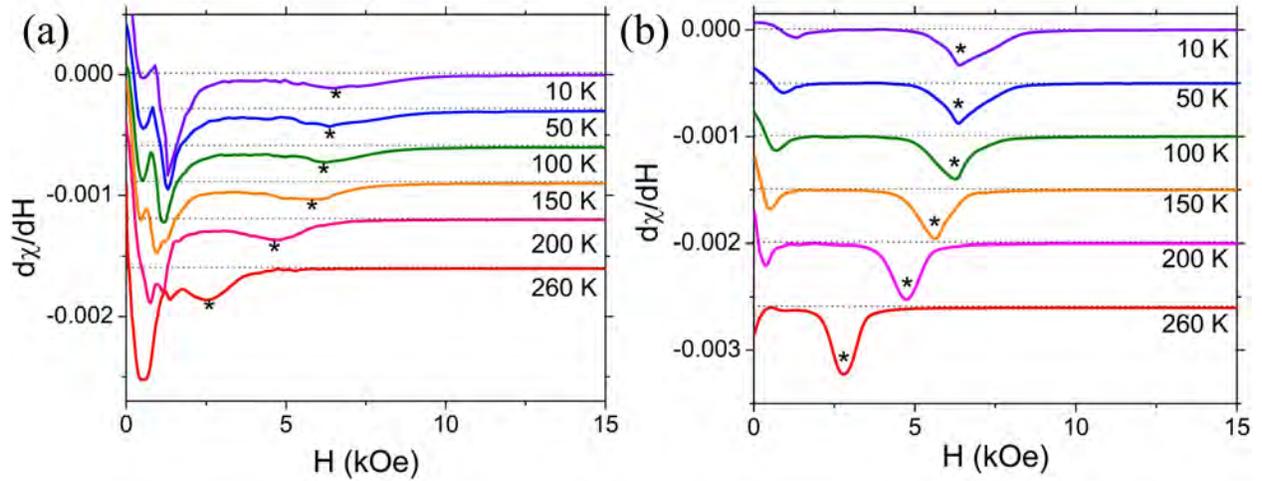

**Figure S10.** Second derivative ($d\chi/dH$) data calculated from the raw $M$ vs. $H$ curves, offset to show different temperature, and presented after filtering the data for (a) 35 nm and (b) 500 nm. The asterisks (*) track the transition into the field polarized state.

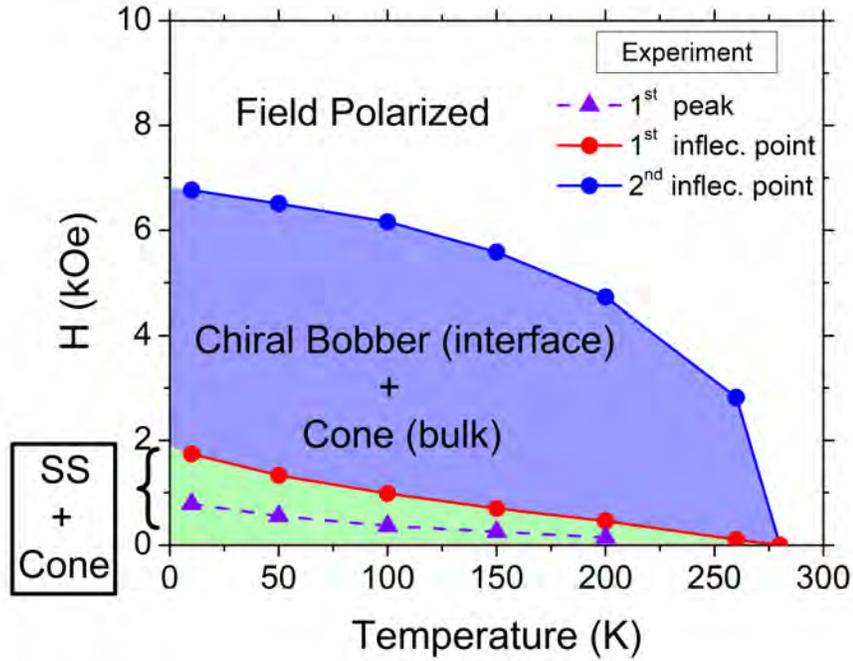

**Figure S11.** *H-T* phase diagram for a 500 nm epitaxial FeGe film showing the wide magnetic field range where the chiral bobber phase is stable similar to what is shown in the micromagnetic phase diagram (see main text Figure 5b).

In Figure S11, we map out the *H-T* magnetic phase diagram for a 500 nm FeGe film ($L/L_D$ ~7) by tracking the peaks and inflection points of $\chi$. By considering the vertical line-cut at $L/L_D = 7$ in Figure 5 in the main text, we identify the phase just below field polarized phase as the chiral bobber / cone phase and the low field phase to be the stacked spiral / cone phase (SS + Cone).

A final point to address is that, in the simulations that follow, we will only consider interfacial Rashba DMI at the FeGe/Si interface but not the FeGe/vacuum interface. The discussion is as follows: at a vacuum interface the effect of mirror symmetry breaking results in "surface twists" [12], which was shown earlier by Rybakov [13] to only be able to produce metastable chiral bobbers. In order to have stable chiral bobbers, it is required to have a sufficiently strong interfacial Rashba DMI (see Figure 5a in the main text). Stable chiral bobbers would not happen at the FeGe/vacuum interface, but could happen at the FeGe/Si interface because the likely presence of strain or interfacial bonding with Si could produce the enhanced spin-orbit coupling and magnetic anisotropy needed to have a sufficiently strong interfacial DMI to stabilize the chiral bobber phase. This is why it is much more likely have to have the chiral bobbers at a single interface (FeGe/Si) rather than both interfaces for our films.

**4. Theory and simulations**

**Micromagnetic simulations**

All micromagnetic simulations for this paper were performed using mumax3 [14] with material parameters for FeGe estimated in [15]: saturation magnetization $M_{sat} = 384$ kA m$^{-2}$, ferromagnetic exchange $2J = A_{ex} = 8.78$ pJ m$^{-1}$, and Dzyaloshinskii-Moriya (DM) exchange $D = D_{bulk} = 1.58$ mJ m$^{-2}$. Important scales that arise are the helical pitch $L_D = J/D$, the characteristic energy density $K_D = D^2/J$, and the characteristic field scale $H_D = K_D/\mu_0 M_{sat}$.

We also include an effective uniaxial anisotropy $K_{eff} = 0$ and $-0.25 K_D$ in Figure 5a and Figure 5b of the main text (negative sign indicates easy-plane anisotropy). The axis for anisotropy is perpendicular to the film plane and aligned with the magnetic field. The value $K_{eff} = 0$ is used as a reference to compare our results with interfacial DMI to previous simulations where interfacial DMI is not included [12]. Below we give a detailed explanation for how we estimate $K_{eff}$ in our thin films. In addition to known material parameters for FeGe we include spatially varying interfacial DM exchange with strength $D_{int}(z)$ where the z-axis is parallel to the film normal. To include spatially varying DMI we had to modify mumax3. The bulk and interfacial types of DM exchange are defined by the energy functionals

$$E_{bulk} = -D_{bulk} \boldsymbol{m} \cdot (\nabla \times \boldsymbol{m})$$

$$E_{int} = -D_{int} \boldsymbol{m} \cdot ([\hat{\boldsymbol{z}} \times \nabla] \times \boldsymbol{m})$$

The film axis is identified as the $\hat{z}$ axis in our simulations. Interfacial DMI is allowed by broken mirror inversion symmetry at the interface of a film; however, previous calculations for thin films have only included $D_{bulk}$ and have ignored $D_{int}$. The value of interfacial DMI is determined by the microscopic physics of the interface and can be very different at a vacuum interface compared with a FeGe/Si(111) interface. The additional surface energy coming from $D_{int}$ is crucial to stabilize the chiral bobber phase on that surface. Note that our simulations do not ignore surface twist effects that arise from free boundary conditions at the interface, and these effects work along with interfacial DMI to enhance the stability of surface spin textures.

**Zero-temperature equilibrium phase diagram**

To obtain the phase diagrams in Figure 5 we use a variational procedure where a set of initial configurations are evolved to a local minimum using conjugate gradient methods. The initial configurations we consider are: conical, helical, stacked spiral, skyrmion tube, and chiral bobber. We also vary system size to find the optimal size for each texture which has interesting implications for the susceptibility discussed below. After all local minima are found we accept the configuration with the lowest energy as the ground state. Our minimization procedure is repeated for different values of system thickness and applied magnetic field to create the phase diagrams shown in Figure 5.

In contrast to previous results in the absence of interfacial DMI [16], we find a wide region of stability for the chiral bobber phase in Figure 5 where we have included interfacial DM exchange near one surface of the simulated film. The interfacial DM exchange $D_{int}(z)$ is confined to a 20 nm layer near one film surface with a magnitude equal to half of the magnitude of bulk DM exchange ($D_{int}$ = 0.79 mJm$^{-2}$) for Figure 5a, and for Figure 5b the penetration of interfacial DMI is increased to 35 nm with a magnitude of $D_{int}$ = 1.19 mJ m$^{-2}$. The increased stability of chiral bobbers in the presence of interfacial DM exchange has an intuitive explanation as follows. Surface spin textures are stabilized by two effects that arise from broken mirror symmetry at the surface, and these effects work together. Previous studies have considered only the effect of free boundary conditions at the interface and find that surface twists increase the stability of stacked spiral, skyrmion, and helical phases [12]. The broken mirror symmetry at a material interface allows for interfacial DMI in addition to the surface twist effect. Broken mirror symmetry allows for both surface twist effects and interfacial DMI. By including both surface twist effects and interfacial DM exchange, surface phases are enhanced relative to conical and field polarized phases, and the effects are compounded since both are compatible with broken mirror symmetry. With the combined effects of surface twist and interfacial DMI the chiral bobber phase becomes stable shown in Figure 5. This is in contrast with previous results where only the surface twist is considered and the cone phase is always more stable than chiral bobbers [12].

The impact of interfacial DMI on surface spin textures is dramatic. Previous calculations have shown that surface twists induced by boundary conditions can lead to stable surface spin textures. Here we have shown that interfacial DMI further increases the stability of surface spin textures, and with sufficiently large interfacial DMI the cone phase is completely removed from the phase diagram and it is replaced by surface spin textures (where the bulk is in a cone phase). Note that the effect of surface spin textures is difficult to observe in measurements of bulk samples with thickness $L \gg L_D$ since surface spin textures contribute $L_D/L$ to any thermodynamic quantity; however, surface modulations are expected even in large samples with $L \gg L_D$ and can be observed with surface sensitive experiments like MFM or spin-polarized STM.

**Estimate for anisotropy of FeGe thin-films**

We use the anisotropy $K_{eff} = K_{u1} + K_M$ which includes both intrinsic easy-axis anisotropy $K_{u1}$ and the easy-plane anisotropy arising from dipolar fields, $K_M = -\mu_0(M_{sat})^2/2$. We have used our experimental magnetization curves to estimate $K_{eff}$ as follows. From our simulations we obtain the upper critical field ($H_{c2}$) for 35 nm and 500 nm FeGe films as a function of anisotropy. For the 35 nm film the upper critical field occurs at a transition between a skyrmion crystal and a field polarized state. We find an empirical form for this phase boundary from our simulations

$$H_{SkX \leftrightarrow FM}(K_{eff}) = H^* - \alpha K_{eff} M_{sat}.$$

From our simulations we find $H^* = 1.1 H_D$ and $\alpha = 1.0$. Next we use our experimental value for $H_{SkX \leftrightarrow FM} = 1.51 H_D$ to determine $K_{eff} = -0.30 K_D$ (easy-plane) in our 35 nm film.

For the 500 nm film the upper critical field occurs at a transition between a cone-like phase with surface modulations (chiral bobbers on the surface) and a field polarized phase. The presence of surface modulations in a 500 nm film does not have a significant impact on the phase boundary since the effect of surface textures is of order $L/L_D$. Thus, the value of $H_{c2}$ for a 500 nm film is well-approximated by the bulk value of $H_{c2}$ for a cone phase which is

$$H_{cone \leftrightarrow FM}(K_{eff}) = H_D - 2 K_{eff} M_{sat}$$

with no fitting parameters. Using the value of $H_{cone \leftrightarrow FM} = 1.49 H_D$ for our 500nm film we find an anisotropy value of $K_{eff} = -0.29 K_D$.

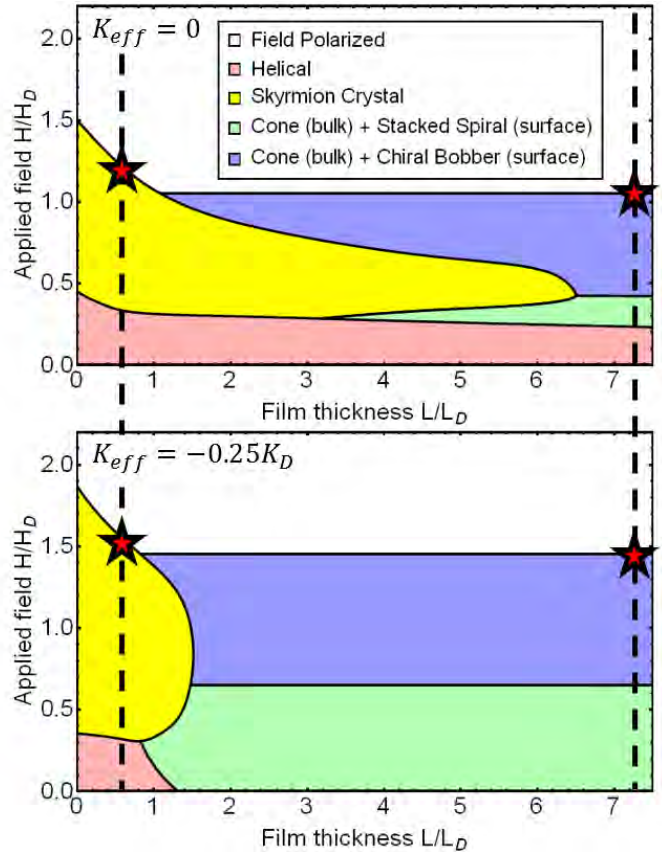

**Figure S12.** The simulations used to produce the *H-L* phase diagram in the main text is also used to determine the anisotropy. The stars label critical field values for a 35 nm film and a 500 nm film in the two cases of no anisotropy and easy-plane anisotropy. For 35 nm films the critical field is a transition from skyrmion crystal to field polarized phase. For 500 nm films the critical field is a transition from a chiral bobber phase to field polarized phase. The critical field value as a function of anisotropy was used to extract the anisotropy value for our films.

**Magnetic susceptibility from equilibrium phase diagram**

The ground state data obtained from our minimization procedure does not show a trend of decreasing susceptibility at high magnetic fields. Especially, the transition between skyrmion crystal and field polarized phases is interesting. Near the field polarized transition skyrmions are predicted to separate rapidly with skyrmion lattice spacing diverging at the transition [17].

Diverging skyrmion lattice spacing leads to increasing susceptibility with a peak at the continuous transition between skyrmion lattice and field polarized phases. To the best of our knowledge diverging skyrmion lattice spacing has never been observed in experiments and our LTEM data show that skyrmion lattice spacing remains roughly constant across the transition. Recent experiments on thin plates of FeGe show similar behavior with skyrmion lattice spacing that is nearly constant as a function of magnetic field [18]. Instead, tightening of the skyrmion core is observed using electron holography, and the field polarized transition is reached through the removal of individual skyrmions [18].

**Confined skyrmion**

In contrast to an equilibrium skyrmion lattice with diverging lattice spacing at the field polarized transition, a single confined skyrmion exhibits decreasing susceptibility near the field polarized transition. To establish this, we perform micromagnetic calculations as follows. With zero applied field we find a local minimum skyrmion configuration using conjugate gradient minimization. Next we increase the field in small steps (0.001T) and perform conjugate gradient minimization for each field value. Throughout the procedure the system is in a local minimum with a single skyrmion, even above the equilibrium transition field. At high fields (above the equilibrium critical field) the local energy minimum becomes too shallow and the system becomes polarized through a first order transition. Susceptibility vs. magnetic field for data obtained with this procedure shows a trend of decreasing susceptibility at high fields. At the transition to the field polarized phase there is a jump in susceptibility that is not observed in our experimental data. This peak is a result of the first order nature of the transition from a single confined skyrmion to the field polarized phase. To understand how this first order transition might be resolved in experimental data we consider disorder broadening which can be captured by a distribution of critical fields.

**Distribution of critical fields**

A second component necessary to explain the observed $d\chi/dH$ trend is broadening of the transition which would occur if the critical field changes between regions of the sample. A distribution of critical fields can arise if uniaxial anisotropy, film thickness, chiral domain size, or some other material parameter, changes on length scales large compared to the skyrmion spacing. Critical field as a function of anisotropy has been calculated previously [19]. As a toy model we can assume a simple analytic form for the magnetization as a function of critical field, in particular, we choose a form that approximates the magnetization data for a confined skyrmion, and then we scale the field axis to match the critical field value. Next, we assume a Gaussian distribution of critical fields. The average magnetization for this toy model preserves the decreasing susceptibility found for the confined skyrmion and it broadens out the first order transition. By combining effects of skyrmion confinement and a distribution of critical fields we can produce magnetization curves in qualitative agreement with the experimental data. However, without a known distribution of material parameters this model contains many fitting parameters and it is not useful to make quantitative comparisons with our experimental data. Future experiments using local probes, e.g., nanoscale FMR, could uncover a distribution of material parameters in thin films of FeGe and help to establish this explanation for the observed trend in *dχ/dH*.